\documentclass[aps,amssymb,amsmath,prd,twocolumn,
showpacs,preprintnumbers,superscriptaddress,nofootinbib,floatfix]{revtex4-1}
\usepackage{graphicx}
\usepackage{bm}
\usepackage{tabularx}
\usepackage{mathrsfs}
\usepackage{latexsym}
\usepackage{float}
\usepackage{color}

\usepackage[normalem]{ulem}
\usepackage{dcolumn}
\usepackage[colorlinks=true,citecolor=blue,urlcolor=blue]{hyperref}
\usepackage[usenames,dvipsnames]{xcolor}
\usepackage[caption=false]{subfig}
\usepackage{slashed}

\newcommand{\be}{\begin{equation}}
\newcommand{\ee}{\end{equation}}
\newcommand{\beq}{\begin{eqnarray}}
\newcommand{\eeq}{\end{eqnarray}}

\newcommand{\bv}{Brunt-V\"ais\"al\"a}

\begin{document}
\title{Temperature effects on core g-modes of neutron stars} 

\author{Nicholas Lozano}
\email{nick.lozano@student.csulb.edu}
\author{Vinh Tran} 
\email{vinh.tran02@student.csulb.edu} 
\author{Prashanth Jaikumar}
\email{prashanth.jaikumar@csulb.edu}
\affiliation{Department of Physics and Astronomy, California State University Long Beach, Long Beach, California~90840, USA}

\date{\today}

\begin{abstract}
Neutron stars provide a unique physical laboratory to study the properties of matter at high density. We study a diagnostic of the composition of high-density matter, namely, g-mode oscillations, which are driven by buoyancy forces. These oscillations can be excited by tidal forces and couple to gravitational waves. We extend prior results for the g-mode spectrum of cold neutron star matter to temperatures that are expected to be achieved in neutron star mergers using a parameterization for finite-temperature effects recently proposed by Raithel, \"Ozel and Psaltis. We find that the g-modes of canonical mass neutron stars ($\approx$1.4$M_{\odot}$) are suppressed at high temperature, and core $g$-modes are supported only in the most massive ($\geq $2$M_{\odot}$) of hot neutron stars.
\end{abstract}

\maketitle


The study of pulsations of stars has a rich history, beginning with the observations of variable stars (eg., Cepheids and RR Lyrae) and continuing through the last century with treatises by Ledoux \& Walraven~\cite{1958HDP....51..353L}, Cox~\cite{1980tsp..book.....C}, Unno~\cite{1989nos..book.....U} and others~\cite{1995ARA&A..33...75G}. Helioseismology and asteroseismology are considered mature fields that have provided important insight into the inner structures of the star as well as the mechanisms that power their variability. Compact stars such as neutron stars and black holes can support pulsation modes that couple to gravitational waves, providing a fingerprint of the equation of state (for neutron stars)~\cite{2001MNRAS.320..307K}, or global properties such as mass and spin rate (for black holes)~\cite{1999LRR.....2....2K}. Within the context of neutron stars, which is the focus of this work, of particular interest is the $g$-mode: a type of non-radial fluid oscillation that is driven by buoyancy forces arising from temperature and composition-related gradients in the star~\cite{1992ApJ...395..240R}. Strong tidal effects during the inspiral phase of neutron star mergers could excite $g$-modes that are potentially observable with the next generation of gravitational wave detectors. 

While many studies of the $g$-mode abound in the literature (e.g.,~\cite{2021PhRvD.103l3009J} and references therein), most hew to the discussion of discontinuity $g$-modes at zero temperature, for e.g.~\cite{2020PhRvD.101l3029T, 2001PhRvD..65b4010S}. In contrast, fewer works have studied the composition $g$-modes pioneered in the work of Reisenegger and Goldreich~\cite{1992ApJ...395..240R} or the effect of high temperatures (${\cal O}$(MeV)) on them. This work is a focused study of how temperature affects the composition $g$-mode through lifting of degeneracy of neutron star matter. In general, the detection of $g$-modes through gravitational wave astronomy could play an important role in uncovering the composition of neutron stars in the near future, including various forms of exotic matter in them.

For the benefit of the busy reader or non-expert, we summarize the main conclusion of this work up front. We find that $g$-modes of hot neutron stars are very different in frequency than those of cold neutron stars, and that the former are suppressed for canonical mass neutron stars due to thermal effects that lead to an increase in the equilibrium sound speed at typical core densities. A detailed discussion is provided in Sec 4.4 and Sec 5. The paper is organized as follows: we begin with a description of the core $g$-mode in neutron stars and its relation to the equilibrium and adiabatic sound speeds in the star through the principal equations used to compute the $g$-mode frequencies within the relativistic Cowling approximation~\cite{2018JCAP...12..031R}, which neglects the associated perturbations of the background metric. In order to extend our analysis of $g$-mode frequencies to high temperatures, we find it convenient to utilize an analytic parameterization for temperature effects on the equation of state for hot and dense matter developed by Raithel, \"Ozel and Psaltis~\cite{2019ApJ...875...12R,2021ApJ...915...73R} and applied to simulations of neutron star mergers. The microscopic physics underlying this parameterization is general enough to apply to most nuclear equation of state, and so we choose two EOS, namely, the Akmal-Pandharipande-Ravenhall (APR) equation of state~\cite{1998PhRvC..58.1804A} and the Zhao-Lattimer (ZL) equation of state~\cite{ZL} for our study. Finally, we interpret our results for the $g$-modes of hot neutron stars and motivate our conclusions. 
\section{$g$-modes and sound speeds in neutron stars}

In general, the oscillatory displacement of a fluid element in a spherically symmetric star is represented by a vector field ${\vec{\xi}}^{nlm}(\vec{r}){\rm e}^{-i\omega t}$ with $n,l$ and $m$ denoting the radial, azimuthal and magnetic mode indices~\cite{2021PhRvD.103l3009J}. To be precise, the frequency $\omega$ also carry subscripts $nlm$ implicitly understood, with degeneracies that are broken in more realistic cases such as with rotation or magnetic fields (not considered here).

For even-parity or spheroidal modes, separation into radial and tangential components yields $\xi_r^{nlm}(\vec{r})$ = $\eta_r^{nl}(r)Y_{lm}(\theta,\phi)$ and $\vec{\xi}_{\perp}^{nlm}(\vec{r})$ = $r\eta_{\perp}^{nl}(r)\nabla_{\perp}Y_{lm}(\theta,\phi)$, respectively, where $Y_{lm}(\theta,\phi)$ are the spherical harmonics~\cite{1967ApJ...149..591T}. From the perturbed continuity equation for the fluid, the tangential function $\eta_{\perp}$ can be traded for fluid variables as $\delta p/\epsilon$ = $\omega^2r\eta_{\perp}(r)Y_{lm}(\theta,\phi){\rm e}^{-i\omega t}$, where  $\delta p$ is the corresponding local (Eulerian) pressure perturbation and $\epsilon$ the local energy density. Within the relativistic Cowling approximation\footnote{The Cowling approximation neglects the back reaction of the gravitational potential and reduces the number of equations we have to solve. While this approximation is not strictly consistent with our fully general relativistic (GR) treatment of the equilibrium structure of the star, it does not change our conclusions qualitatively or even quantitatively that much, since this approximation is accurate for $g$-mode frequencies at the few \% level~\cite{Grig, 2017PhRvC..95b5808F, Zhao:2022toc}.}, the equations of motion to be solved to determine the frequency of a particular mode are (in $c$=1 units)~\cite{1983ApJ...268..837M,1992ApJ...395..240R,2014MNRAS.442L..90K}
\begin{align}
-\frac{1}{\mathrm{e}^{\lambda / 2} r^{2}} \frac{\partial}{\partial r}\left[\mathrm{e}^{\lambda / 2} r^{2} \xi_{r}\right]+\frac{l(l+1) \mathrm{e}^{\nu}}{r^{2} \omega^{2}} \frac{\delta p}{p+\epsilon} 
-\frac{\Delta p}{\gamma p} = 0  \nonumber \\ 
\frac{\partial \delta p}{\partial r}+g\left(1+\frac{1}{c_{\mathrm{s}}^{2}}\right) \delta p+\mathrm{e}^{\lambda-\nu} h
\left(N^{2}-\omega^{2}\right) \xi_{r} = 0 \,, 
\label{oscr}
\end{align}
where enthalpy $h=p+\epsilon$, $\gamma$ the adiabatic index of the fluid, and we have suppressed the indices on $\omega$ and $\xi$. The Lagrangian variation of the pressure enters as $\Delta p$, and is related to the Eulerian variation $\delta p$ through the operator relation $\Delta \equiv \delta + \xi\cdot\nabla$. The symbol $c_s$ denotes the adiabatic sound speed, the square of which is given as $c_s^2=\gamma_1 p/(\mu_n n_B)$ where $\mu_n$ is the neutron chemical potential\footnote{In beta-equilibrated charge neutral neutron star matter, the neutron chemical potential is sufficient to determine all other chemical potentials.}, $n_B$ the local baryon density and $\gamma_1{=}(n_B/p)\partial p(n_B,Y_p)/\partial n_B$ is the adiabatic index. The advantage of employing the analytic parameterization of the finite temperature EOS which separates out the dependence on $n_B$ and $Y_p$ is that the partial derivative $\partial p(n_B,Y_p)/\partial n_B$, and hence the adiabatic sound speed at any density can be computed analytically. On the other hand, the equilibrium sound speed squared $c_e^2=(dp/d\epsilon)$ is a total derivative that can be evaluated numerically from the tabulated EOS. Both sound speeds enter through the \bv\, frequency ($N$) which is given by 

\begin{equation}
\label{bvf}
    {\it N}^{2}\equiv g^{2}\Big ( \frac{1}{c_{e}^{2}}-\frac{1}{c_{s}^{2}} \Big ){\rm e}^{\nu-\lambda} \,,
\end{equation}

where the local gravitational field $g$=-$\nabla\phi$=-$\nabla p/h$, $\nu(r)$ and $\lambda(r)$ are metric functions of the unperturbed star which feature in the Schwarzschild {\it interior} metric. Eq.(\ref{oscr}) can be analyzed in the short-wavelength limit ($kr\gg 1$) where the local dispersion relation has two distinct branches, with the lower frequency branch corresponding to the $g$-modes. The local $g$-mode frequency is then $\omega^2\propto {\rm e}^{\lambda}N^2$~\cite{1983ApJ...268..837M}, highlighting the importance of the two sound speeds (in particular, the difference of their inverse squares, as in Eq.(\ref{bvf})). The global g-mode frequency is constant for a given stellar configuration (see Fig.\ref{G-mode}) and can be thought of as an average of the local g-modes (although it is still sensitive to phase transitions).

In this work, we study the fundamental $g$-mode with $n$ = 1 and fix the mode's multipolarity at $l$ = 2. This is because the $l=2$ mode is quadrupolar in nature, and can couple to gravitational waves. Higher $l$ values (octupole and higher) are generally weaker than the quadrupole. The reason to study the $n=1$ (fundamental) $g$-mode is that the local dispersion relation for $g$-modes $\omega^2\propto 1/k^2$ implies that the $n=1$ excitation has the highest frequency, whereas higher values of $n$ are known to have a smaller amplitude of excitation and a weaker tidal coupling coefficient~\cite{2021PhRvD.104l3032C}. For the non-rotating stars we consider here, solutions are degenerate in $m$. Note that our definition of the ``fundamental" mode refers to the lowest radial order of the $g$-mode which also has the highest frequency. This should not be confused with the qualitatively different $f$-mode which is also referred to sometimes as the fundamental mode. Furthermore, overtones with lower frequency exist, but we do not perform any computations with them here, since the fundamental $g$-mode is likely the most strongly excited during a tidal perturbation and is within the sensitivity range of current generation of gravitational wave (GW) detectors{{~\cite{1994MNRAS.270..611L,2022arXiv220403037Z}}}.

The system of equations in Eq.~(\ref{oscr}) cannot be solved analytically even with a simple model of a neutron star. Our aim will be to solve this numerically as an eigenvalue system for the $g$-mode frequency $\omega$. Physically, the solution to this system of equations, under the boundary conditions $\Delta p=0$ at the surface and $\xi_r,\,\delta p/\epsilon$ regular at the center, only exists for discrete values of the mode frequency $\omega$. These values represent the $g$-mode spectrum for a chosen stellar model. 
Because we have employed the Cowling approximation and ignored the perturbations of the metric that must accompany fluid perturbations, we cannot compute the imaginary part of the eigenfrequency (damping time) of the $g$-mode\footnote{The damping time of $g$-modes due to viscosity and gravitational wave emission, estimated  in some works~\cite{1999MNRAS.307.1001L,2020ApJ...904..187W}, suggests that the $g$-mode can become secularly unstable for temperatures $10^8~{\rm K}<T<10^9~{\rm K}$ for rotational speeds exceeding twice the $g$-mode frequency of a static star.}. 

The effect of temperature on the $g$-mode in degenerate matter inside neutron stars is relatively unexplored. Fu et al.~\cite{FBL} calculated the time-evolution of the $g$-mode spectrum of a newly born neutron star/strange quark star as it cools after formation in a supernova explosion, within a specific model (the relativistic mean-field/MIT Bag model respectively). Our purpose in this work is to explore the behavior of $g$-modes with increasing temperature (decreasing degeneracy) in an evolved neutron star. For this exercise, we employ two parameterized models for a purely nucleonic finite-temperature equation of state that satisfy current observational constraints and are described in more detail below.
\section{Finite temperature model for the neutron star equation of state}
References~\cite{2019ApJ...875...12R,2021ApJ...915...73R} introduced a finite-temperature model (henceforth, ROP) in parameterized form that captures thermal effects in $npe$ matter at arbitrary density, temperature and proton fraction, and can describe a wide range of microscopic realistic EOS. In particular, the model can describe thermal effects on degeneracy and is therefore more widely applicable to the temperature and density regimes relevant to neutron star mergers than the ideal fluid approximation. The authors also tabulate fit parameters for different classes of thermal EOS, such as the single nucleus approach of Lattimer \& Swesty~\cite{1991NuPhA.535..331L}, Shen et al.~\cite{1998NuPhA.637..435S} as well as nuclear statistical equilibrium, e.g., Hempel \& Schaffner-Bielich~\cite{2010NuPhA.837..210H} and Steiner et al~\cite{2013ApJ...774...17S}\footnote{Not all fit parameters in the ROP model are consistent with the combined set of experimental constraints~\cite{2013ApJ...771...51L} or with error bands from chiral effective field theory~\cite{2013PhRvC..88b5802K,2021PhRvC.103d5808D} but a subset of parameters always are, with a reasonable compromise to allow for a wide range of EOS in their initial study.}. 

The relevant aspects of the ROP model for this work are the various thermal contributions to the energy and pressure of neutron star matter, which are all expressed in analytic (exact or parameterized) form, with explicit dependence on temperature, density and compositional dependence (in this case, proton fraction only), facilitating a computation of the equilibrium and adiabatic sound speed required for the finite temperature $g$-mode. Thermal contributions come from the relativistic species (photons and leptons only; neutrinos are ignored), the ideal gas component (non-degenerate nucleons) and a thermal-degenerate component calculated to leading order ($\propto T^2$). Each of these three components dominates in different density regimes, from low-density to high-density respectively, which can be approximated by a smoothed expression that recovers the dominant contribution to the energy and pressure of the strongly interacting matter. In the ROP prescription, the Dirac effective mass of the nucleon (neutron or proton) is also parameterized by a power-law form, which, at the expense of introducing additional fit parameters, reproduces the behavior of the effective mass in the microscopic theory at densities and temperatures relevant to the core of neutron stars~\cite{2015PhRvC..92b5801C,2015AnPhy.363..533C}. The nucleon's effective mass has been shown to be important for other oscillation modes, such as $f$-modes~\cite{2021PhRvC.103c5810P}. Thus, the ROP model is well suited to a computation of the sound speeds and the finite-temperature $g$-mode. For the convenience of the reader, we reproduce the generic expressions for the pressure and energy within the ROP model in the Appendix.

\section{Results}

We present our results for the $g$-mode in this section, beginning with the inputs required for their computation, namely, the EOS, the mass-radius relation, and the sound speeds. 
\begin{figure*}[]
\begin{centering}
\includegraphics[scale=0.45]{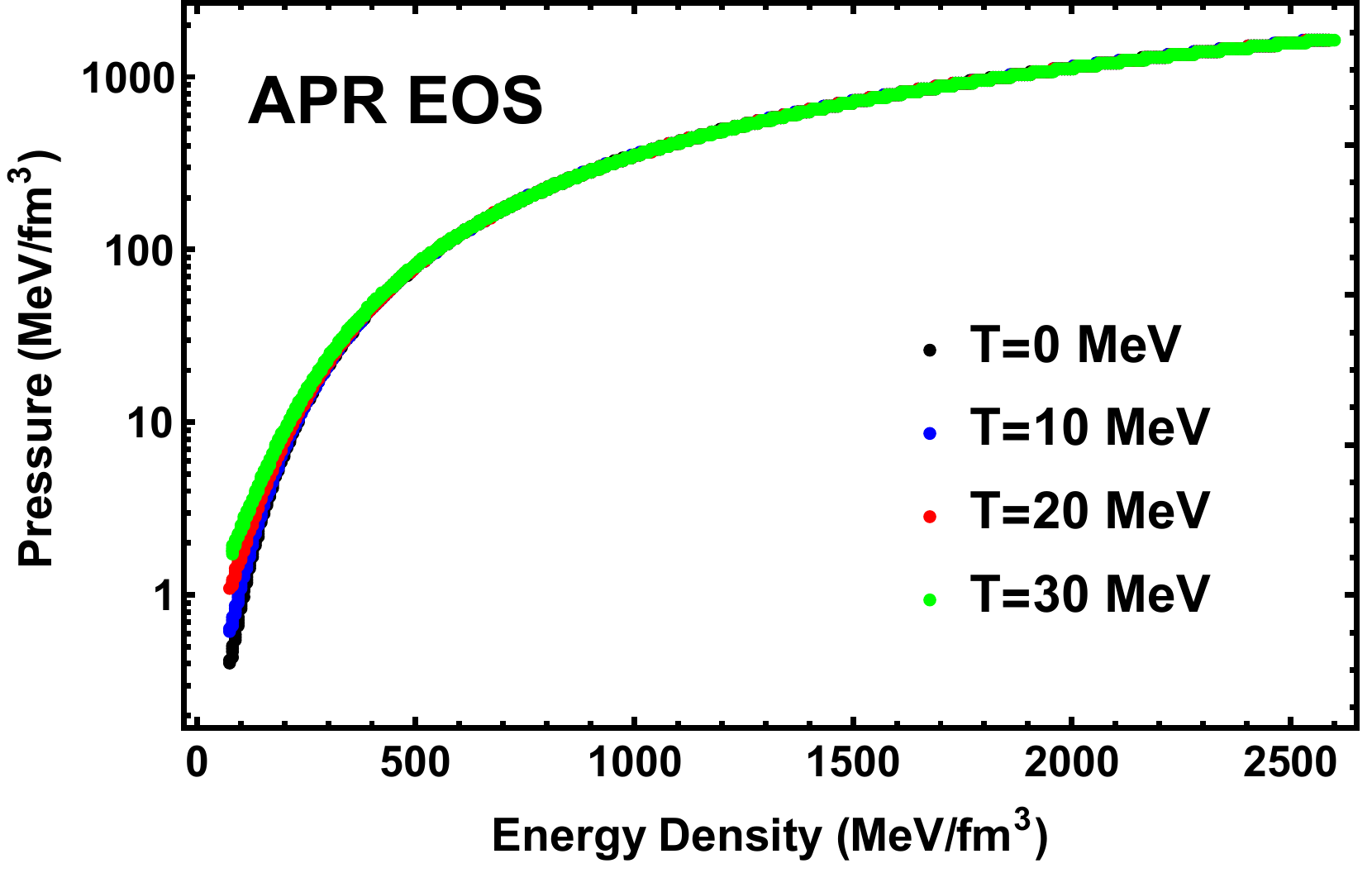} \qquad \includegraphics[scale=0.47]{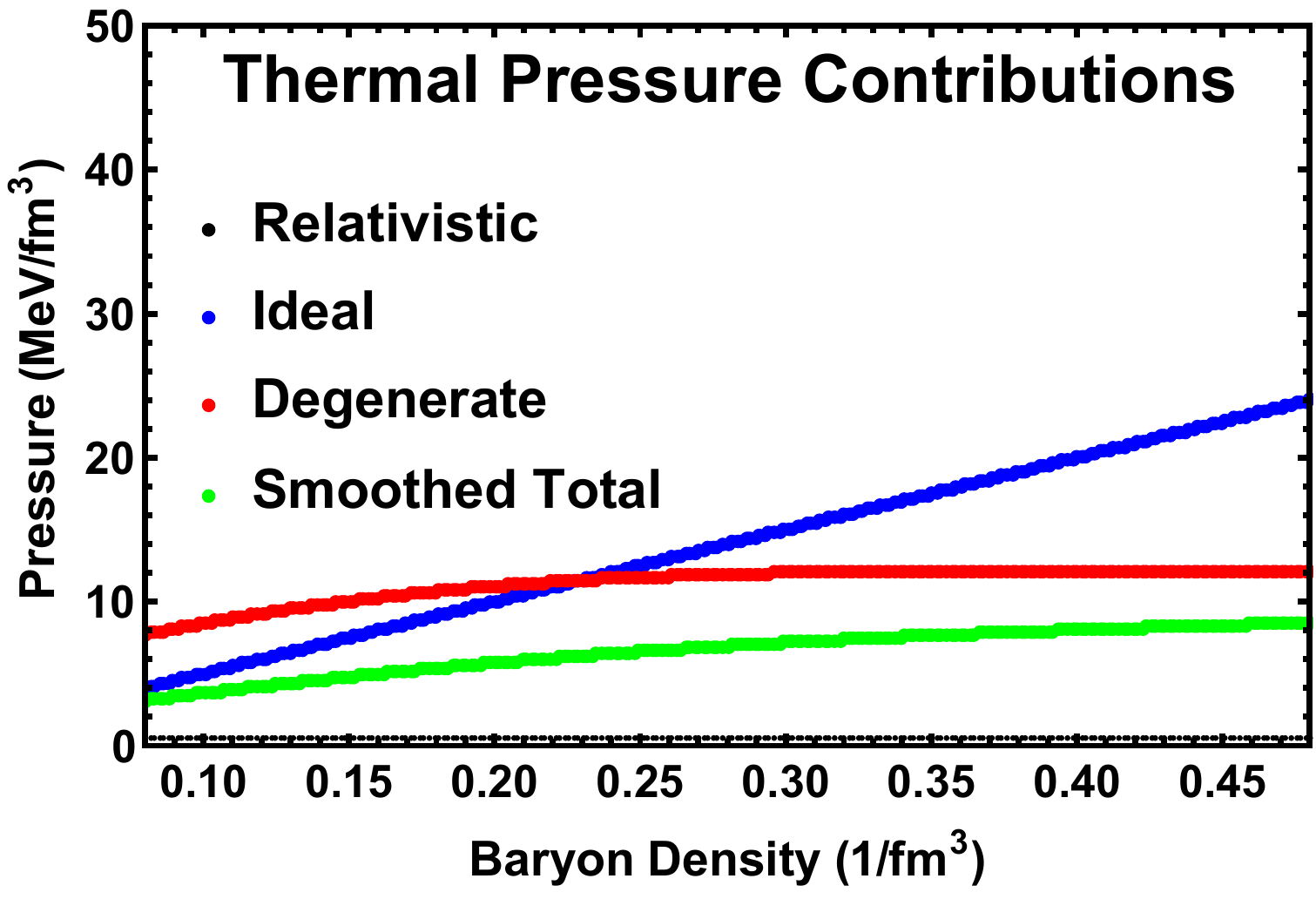}
    \caption{(\textbf{a}) The APR EOS with thermal contributions included as per the parameterizations in~\cite{2019ApJ...875...12R}. (\textbf{b}) Thermal effects are most pronounced in the low-density regime (0.05-3.0) $n_{\rm sat}$ and are thus expected to inflate the radius of the hot neutron star relative to a cold one. This is proportionate to the impact of thermal effects, shown here for a temperature $T$=50 MeV, on lifting the degeneracy of dense matter. Corresponding results for the ZL EOS are qualitatively similar.}
    \label{EOS relation}
\end{centering}
\end{figure*}

\subsection{Pressure and Energy}
We choose two representative nuclear models for the EOS of dense matter in the core of the neutron star. One is the APR EOS~\cite{1998PhRvC..58.1804A}. To construct this cold EOS for use in Eq.(\ref{EOS1}) \& (\ref{EOS2}), we fit the parabolic approximation 
\begin{align} 
E(n,Y_p,T{=}0) &= E_B(n,Y_p{=}1/2,T{=}0)\\
&\qquad +(1{-}2Y_p)^2E_{\rm sym}(n,T{=}0) \nonumber 
\end{align} 
to the results of~\cite{1998PhRvC..58.1804A} with 
\begin{align}
E_B(n,Y_p{=}1/2,T{=}0) = E_0(u{-}2{-}\delta)/(1{+}u\,\delta)
\end{align} 
with $u = n/n_{\rm sat}$ ($n_{\rm sat}$ = 0.154 fm$^{-3}$ is the saturation density), $E_0 = - 16$ MeV the binding energy of uniform nuclear matter and $\delta$ a parameter that is fixed by choosing the value of the compressibility $\kappa$ of nuclear matter at saturation\footnote{ Compressiblity $\kappa$= 18$E_0/(1+\delta)$ for the fitted APR EOS}. It is easy to check that the pressure of symmetric matter vanishes at saturation. The fit for $E_{\rm sym}(n,T{=}0)$ is given by Eq.(14) of~\cite{2019ApJ...875...12R} and the parameters required for this are the symmetry energy at saturation $E_{sym}(n_{sat},T{=}0)$, the density dependence of the potential contribution to the symmetry energy, typically parameterized by a power-law index $\gamma$ and the slope parameter $L$ at saturation. Our parameter values are within the typical range allowed by terrestrial and astrophysical data, as detailed in~\cite{2015PhRvC..91d4601L}. A common nucleon mass is chosen for convenience. With these choices fixed, one can obtain $Y_{p,\beta}(n)$, the proton fraction in charge-neutral $npe$ matter in $\beta$-equilibrium. The compressibility at saturation density $\kappa$ is tuned to satisfy the maximum mass constraint of 2.08 $_{-0.07}^{+0.07}M_{\odot}$ derived from observations of PSR J0740+6620 in~\cite{Fonseca}. All parameter values for the APR EOS are listed in Table~\ref{tab1}.

\begin{table}[b] 
\caption{Parameters of the cold EOS employed in this work.\label{tab1}}
\begin{ruledtabular} 
	\begin{tabular}{ccccc}
		\textbf{EOS}	& \textbf{$\kappa$}	& $E_{\rm sym}(n_{\text{sat}})$ (MeV)	& $L$ (MeV) & $\gamma$ \\
		\hline 
		\hline 
		APR & 254 & 32 & 57.6 & 0.6 \\ 
		ZL & 250 & 31.6 & 60.1 & 0.8 
	\end{tabular}
\end{ruledtabular} 
\end{table}

The second EOS model is that of Zhao and Lattimer~\cite{ZL}. In this case, the data for $E_B(n,Y_p{=}1/2,T{=}0)$ is fitted to the compressibility of nuclear matter, along with a binding energy of -16 MeV and vanishing pressure, all at saturation\footnote{Data for the ZL EOS was obtained in private communication with author of~\cite{2021PhRvD.104l3032C}, while data for $E_{\rm sym}(n,T{=}0)$ is again fit by Eq.(14) of~\cite{2019ApJ...875...12R}. The parameters values $\kappa, E_{sym}(n_{sat}), L, \gamma$ for the ZL EOS are listed in Table~\ref{tab1}.}

For the cold ($T$=0) APR EOS, the dimensionless tidal polarizability $\Lambda_{1.4}^{\rm APR}$=492 while for the more compact ZL EOS, $\Lambda_{1.4}^{\rm ZL}$=357. Both are within the constraint 70 $\leq \Lambda_{1.4} \leq$ 580 obtained using uncorrelated priors on the binary components of the merger event GW170817 obtained by Abbott et al.~\cite{Abbott}, and also within the constraints of De et al.~\cite{De} which used correlated priors. For both EOS, we will treat the crust as a homogeneous fluid so that $c_s$=$c_e$ and no crustal $g$-modes are supported. The assumption of homogeneity is reasonable given that the crust melting temperature is estimated to be well below temperatures of 10 MeV or more that we are considering in our study of finite temperature $g$-modes.

In either case, adding the thermal contributions gives us the complete EOS at any temperature and density relevant to this study; Eqs.(\ref{Full_Energy}) and (\ref{Full_Pressure}). Figure~{\bf\ref{EOS relation}(a)} shows the pressure-energy relation for the APR EOS in $\beta$-equilibrium for varying temperatures including the density regime of relevance to the core of neutron stars, while Figure~{\bf\ref{EOS relation}(b)} shows the relative importance of thermal contributions as a function of density for a fixed temperature. The smoothed pressure is the sum of the relativistic term and the harmonic mean of the ideal and the degenerate terms, and is employed to avoid artificial discontinuities in the calculation of sound speeds. It also has the right limits in that it approaches the ideal gas pressure at low densities and the degenerate pressure at high density. The corresponding results for the ZL EOS are qualitatively similar, hence not shown here.

\subsection{Mass-radius relation}
\begin{figure}[h!]
\centering
    \includegraphics[scale=0.4]{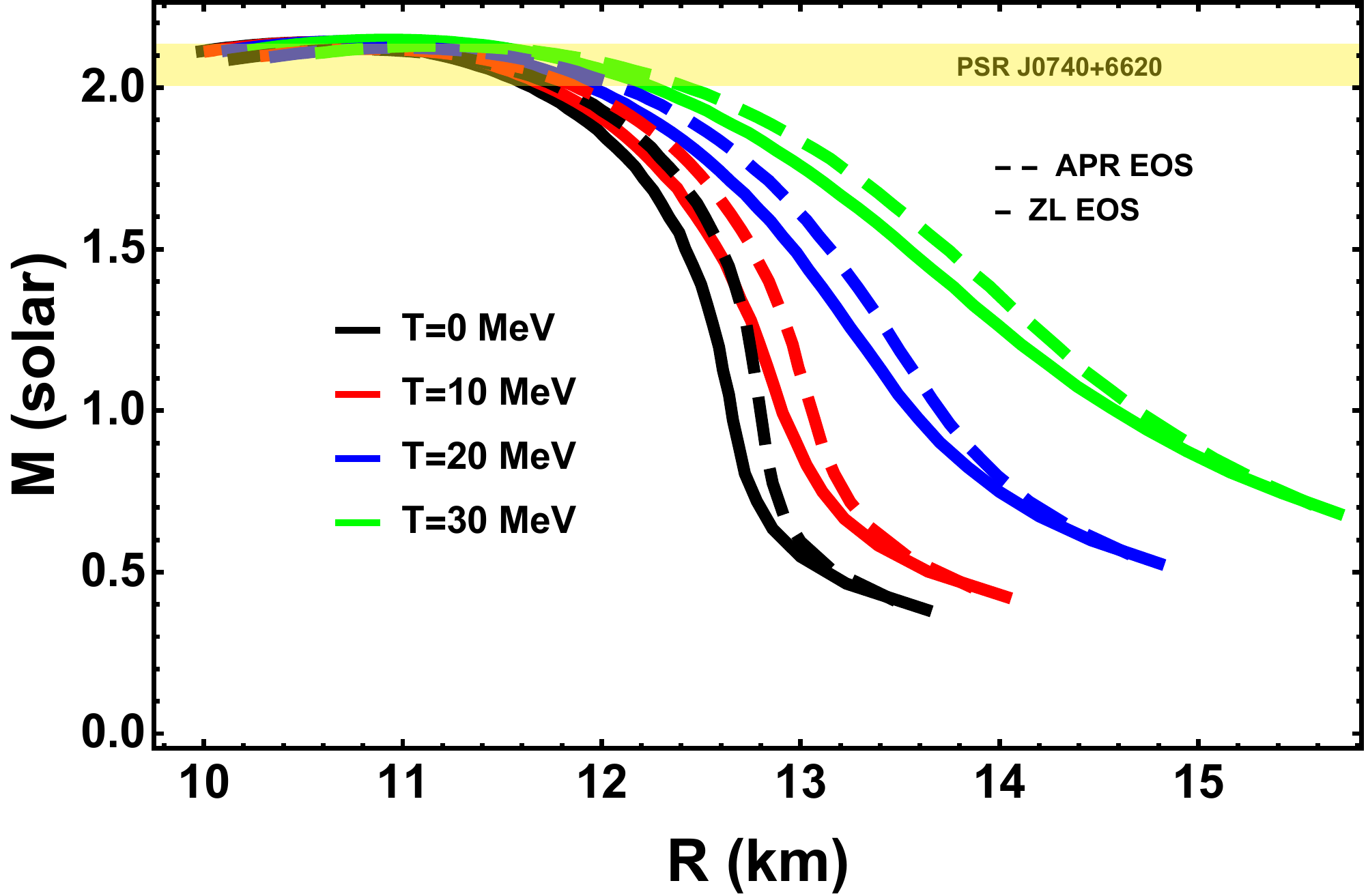}
    \caption{The effects of temperature on a neutron star's structure. As the temperature increases, degeneracy is lifted. This means that the stars will essentially ``puff'' up without gaining more mass.  }
    \label{mass radius relation}
\end{figure}
The mass-radius relations of cold and hot neutron stars described by the parameterized APR EOS and the ZL EOS are shown in Figure~\ref{mass radius relation}. The maximum mass of the $T$=0 neutron star is 2.12 $M_{\odot}$ for the APR EOS and 2.15 $M_{\odot}$ for the ZL EOS, representative of the heavier neutron stars observed to date, giving us a sufficiently wide range to study the $g$-mode dependence on the mass. With increasing temperature ($T$ from 0 to 30 MeV), the maximum mass for the APR EOS changes only slightly to 2.15 $M_{\odot}$, while the size of the star increases, with dramatically larger radii near the upper limit of this range. Such high temperatures can be achieved a few tens of milliseconds after the merger in the core of the remnant, as suggested by results for the thermal profile emerging from merger simulations~\cite{2021PhRvD.104f3016R}. It should be noted that at this stage, the remnant is not in a state of mechanical or chemical equilibrium, so our mass-radius profiles are not directly applicable to such a scenario. However, it does demonstrate that the radial extent of the star is quite sensitive to the treatment of thermal pressure. Including the thermal terms leads to radius inflation and as we note below, affects the sound speeds and the behavior of the $g$-mode frequency.

\subsection{The equilibrium ($c_e$) and adiabatic ($c_s$)  sound speeds}

\begin{figure*}
\begin{centering} 
\includegraphics[scale=0.5]{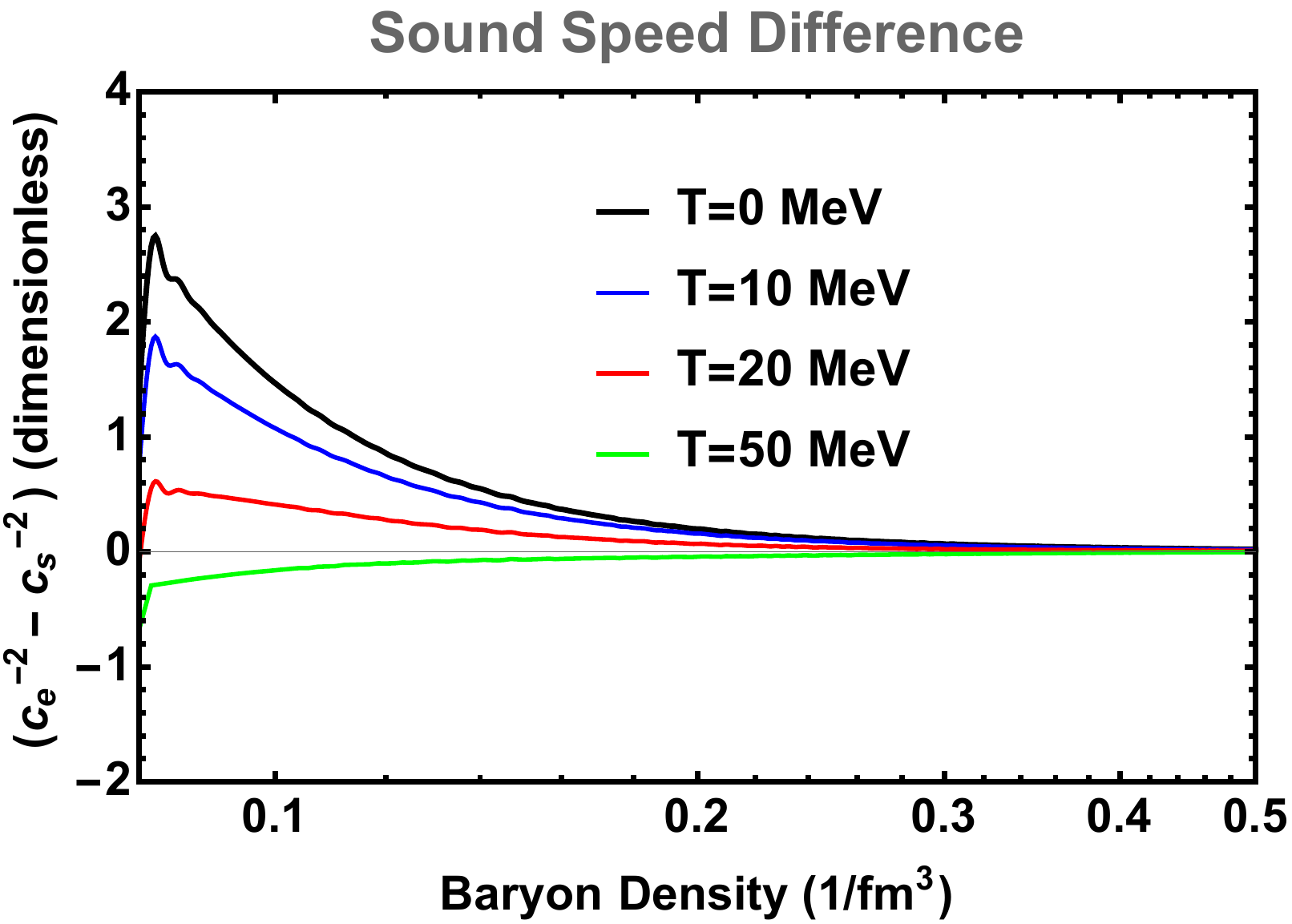} \qquad 
\includegraphics[scale=0.5]{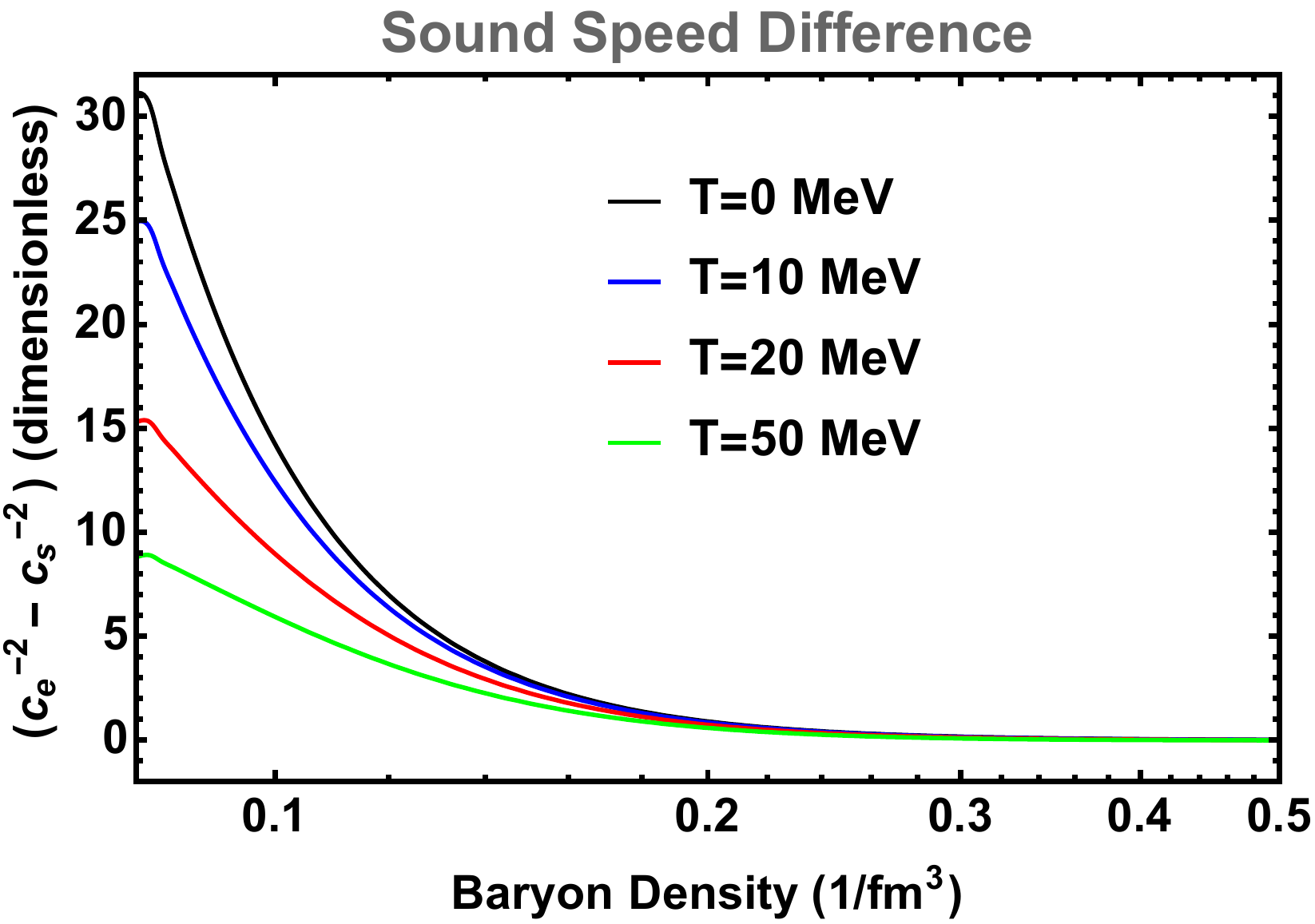}
    \caption{The difference between the inverse of the squares of the equilibrium and adiabatic sound speeds as a function of density in the neutron star core for varying temperatures, for the APR EOS. A positive difference indicates stable $g$-modes. (\textbf{a}) For $Y_p$ the same as that given by the cold EOS (i.e., in $\beta$-equilibrium), at temperatures of 50 MeV or more, no stable $g$-modes are supported . (\textbf{b}) For a fixed, arbitrary $Y_p=0.1$, this difference, though positive at low density eventually turns negative at large density irrespective of the temperature. Interpolation effects at the core-crust boundary ($n_B$=0.08 fm$^{-3}$) cause the peaks in the left panel figure and do not have any physical significance. In practice, the sound speed difference in the core rises smoothly towards the core-crust boundary and drops abruptly to 0 in the crust in accordance with our assumption of a homogeneous fluid crust (see text for additional discussion on this point). The corresponding results for the ZL EOS are qualitatively similar.}
    \label{bv-fig}
\end{centering}
\end{figure*}

The difference between the (inverse squares of) equilibrium and adiabatic sound speeds determines the driving frequency of the $g$-modes and depends on the composition in $\beta$-equilibrium. As pressure and energy density vary with temperature, we find that the sound speed difference also changes strongly with temperature, which affects the $g$-mode frequency. If we assume the proton fraction remains unchanged from the cold EOS value $Y_p$=$Y_{p,\beta}$, we observe from Figure~\ref{bv-fig}{\bf (a)} that with increasing temperature, this sound speed difference gradually decreases and turns negative, switching off the $g$-modes. At low temperatures, the equilibrium sound speed squared ($c_e^2$) is smaller than the adiabatic sound speed squared ($c_s^2$) at all densities of relevance to the neutron star's core. With increasing temperature, $c_e^2$ exceeds $c_s^2$ at a threshold density which is within the realm of a neutron star's core densities. This increase in $c_e^2$ is principally due to the thermal pressure from the ideal gas term at lower densities, and the degeneracy term at higher densities. On average, the sound speed difference $c_e^2-c_s^2$ is most affected by the lifting of the degeneracy. We conclude that composition $g$-modes in neutron stars should be suppressed at high temperatures. Both figures terminate at a lower density of $n_b$=0.08 fm$^{-3}$ (core-crust interface). In principle, the crust can also support $g$-modes and these can be calculated given a particular model of the crust with shells of varying composition and shear moduli~\cite{Finn}, but crustal $g$-modes couple very weakly or not at all to core $g$-modes~\cite{1992ApJ...395..240R}, hence for our purposes, it suffices to treat the crust as a homogeneous fluid with $c_s$=$c_e$ without any $g$-modes there.

As the model in ROP is a good fit to realistic EOS across a range of asymmetries 0.01 $\leq Y_p \leq 0.5$ in $npe$ matter, we also study the behavior of the sound speed difference with baryon density at fixed $Y_p$. Although the fixed $Y_p$ scenario is not realized in neutron stars and does not apply to the $g$-mode perturbations which assume small deviations from $\beta$-equilibrium, we display the results here to demonstrate the marked effects of composition on sound speeds. For example, Figure~\ref{bv-fig}{\bf (b)} shows that $c_e^2 < c_s^2$ at low densities, even up to high temperatures, as matter in the core of a neutron star is, on average more degenerate than a configuration with fixed $Y_p$=0.1 implies. The corresponding results for the ZL EOS are qualitatively similar, hence not shown here.

\subsection{$g$-modes at finite temperature}
\begin{figure}[h!]
    \centering
    \includegraphics[scale=0.35]{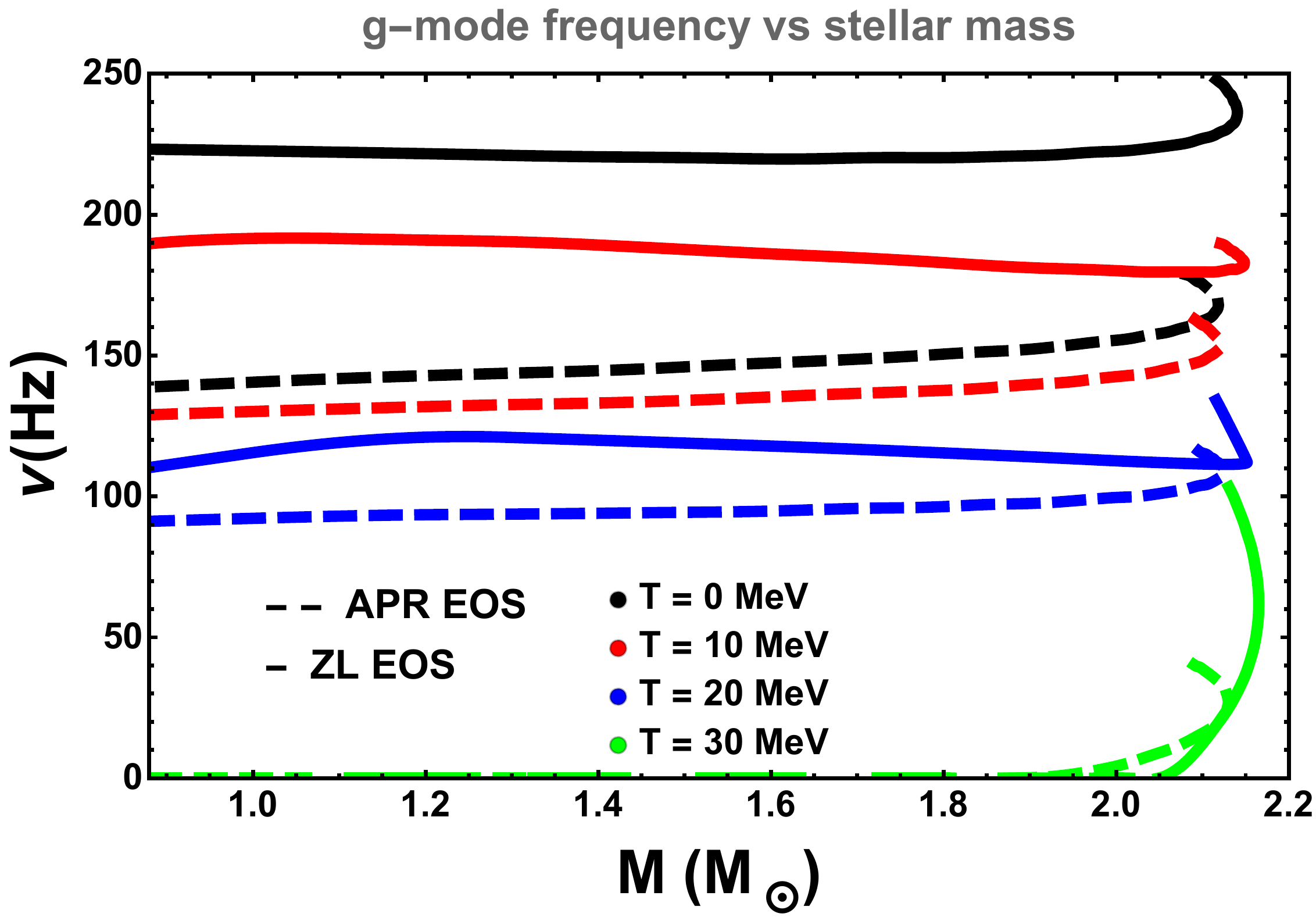}
    \caption{Plots of the g-mode frequencies as a function of the star's mass (in solar masses). As temperature increases, g-mode frequencies are suppressed. At temperatures of 30 MeV, g-modes vanish except near the maximum mass, and eventually these disappear as well with increasing temperature.}
    \label{G-mode}
\end{figure}

For composition-driven $g$-modes of zero-temperature neutron stars, the scale of the characteristic frequency is set by the \bv\, frequency, and is of the order of 100-200 Hz. The $g$-mode frequency increases slowly with stellar mass, until a sharp increase occurs near the maximum mass. The $g$-mode frequency has been found to scale with the central lepton fraction~\cite{Zhao:2022toc}, and is sensitive (in $npe$ matter) to the symmetry energy and its density derivative~\cite{2020ApJ...904..187W}. As Figure \ref{G-mode} shows, these modes are strongly dependent on the star's temperature as well, with g-modes for canonical neutron stars (1.4 $M_{\odot}$) being suppressed at or above temperatures of 30 MeV or so. This is a direct result of an increase in the equilibrium sound speed with increasing temperature. Neutron stars close to the maximum mass of 2$M_{\odot}$ in the APR or ZL EOS model can support stable $g$-modes even at higher temperature, but eventually at temperatures of 50 MeV or more, no stable $g$-modes can be found. 

\section{Discussion}
We studied the effects of temperature on the composition-driven $g$-modes of a neutron star, which could be excited during or in the aftermath of a neutron star merger, and possibly be detected in future gravitational wave detectors. Toward this end, we employed a convenient parameterization of finite temperature effects that accurately reproduces exact results from a variety of microscopic calculations of finite-temperature equations. In particular, the separation of the density and composition dependence of the parameterized form, for both the cold and thermal contributions to the pressure, facilitates the computation of the $g$-mode frequencies. 

Several effects can be seen with temperatures of order 10 MeV or more which characterize hot neutron stars. In general, higher temperatures alter the equation of state, leading to higher energy density and pressure. As a result, the mass-radius relation of the stars changed as a function of temperature (see Figure \ref{mass radius relation}).  For example, with the APR EOS, a star with a mass of 1.4 solar mass at low temperature has a radius of about 12.45 km. At 30 MeV, a star with similar mass has a radius of 13.68 km. The increased temperature lifts the degeneracy and puffs up the star creating an overall slightly less dense neutron star on  average when compared to lower temperatures. 
\par
Additionally, the calculated frequencies of the $g$-mode  decreased as temperature increased (Figure \ref{G-mode}). One observes that at high temperatures $g$-mode oscillation frequencies are suppressed and disappear almost completely beyond 30 MeV, except in the most massive of neutron stars near the maximum mass configuration, due to the stronger effect of degeneracy in the cores of such stars. This suppression is due to specific thermal contributions in the equation of state which makes the equilibrium sound speed larger than the adiabatic sound speed. 
\par
Since the merger of two neutron stars results in a hotter and more massive neutron star than the merging components, one may speculate that $g$-modes can still be marginally supported in the remnant due to its highly degenerate core. Such remnant stars vibrate strongly for a few seconds, so gravitational wave detectors may detect $g$-modes from them, assuming viscosity does not damp the $g$-mode too quickly. The detection of zero-temperature $g$-modes from neutron stars in the pre-merger phase has been studied extensively, and may also be a promising avenue to constrain the composition of cold neutron stars. 
\par
Some caveats to our study are to be noted. The composition $g$-modes introduced by Reisenegger and Goldreich~\cite{1992ApJ...395..240R} are, strictly speaking, excited in non-convective (cold) neutron stars. If temperatures become large enough for convective forces, such $g$-modes can become unstable. Convective forces are not taken into account in this work, nor are temperature gradients. For a varying temperature profile (e.g., assuming a constant entropy/baryon), a more general expression for the adiabatic sound speed such as eq.(56) of ~\cite{Kap} is required. A more detailed investigation of the $g$-mode with a realistic temperature profile and a locally temperature dependent sound speed is much more involved and left to future work. Neutrino trapping is also ignored, but if we nominally include them as a thermal component by setting $f_s=43/8$ above typical trapping temperatures of 10 MeV, we find a systematic reduction of approximately 15-20\% in the $g$-mode frequency. This is significant and implies that neutrinos can affect the $g$-mode frequency in hot neutron stars, a finding worthy of additional study. The inclusion and impact of additional species (such as hyperons or quarks) can all cause a substantial enhancement of the $g$-mode frequency by about 500 Hz, though in each case the reason for the enhancement is different (see, e.g., discussion in~\cite{2021PhRvD.104l3032C}). As the $g$-mode is expected to be swamped by the $f$-mode, it has not been the focus of attention in the early LIGO/VIRGO studies (e.g.,~\cite{2020NatCo..11.2553P}) but the important compositional information carried by the $g$-mode may be teased out in future detectors such as the Cosmic Explorer or the Einstein Telescope. 

\begin{acknowledgments}

Authors acknowledge helpful comments from an anonymous referee.
P.J. is supported by the U.S. National Science Foundation Grant No. PHY-1913693.

\end{acknowledgments}

\appendix
\section[\appendixname~\thesection]{}
\subsection[\appendixname~\thesubsection]{ROP parameterization for the temperature dependent EOS}
A note about units: Following convention, in the governing $g$-mode equations of  Sec. 2, we have set $c$=1 and chosen MeV/fm$^3$ for units of energy density and pressure. For consistency, in this section, $E$ is in MeV and $p$ in MeV/fm$^3$, so one should use as appropriate the conversions MeV.fm = 1/197.33, while measuring $T$ in MeV (1 MeV = 1.16$\times 10^{10}$K) and number densities in fm$^{-3}$. The energy and pressure are decomposed as
   \begin{align}
\label{EOS1}
E_{\text{total}}(n,Y_p,T) = E_{\text{cold}}(n,Y_p)+E_{th}(n,Y_p,T) \,.\\
p_\text{{total}}(n,Y_p,T) = p_\text{{cold}}(n,Y_p)+p_{th}(n,Y_p,T)\,.
\label{EOS2}
\end{align}

where $E_{th} (E_{cold})$ and $p_{th} (p_{cold})$ represent the total respective thermal (cold) components. In Eq.(\ref{EOS1}) \& (\ref{EOS2})

\begin{widetext} 
\begin{multline}
E_{\text{cold}}(n,Y_p) = E\left(n, Y_{p, \beta}, T=0\right)
+E_{\text{sym}}(n, T=0)\left[\left(1-2 Y_{p}\right)^{2}-\left(1-2 Y_{p, \beta}\right)^{2}\right] +  3K\left(Y_{p}^{4 / 3}-Y_{p, \beta}^{4 / 3}\right) n^{1 / 3} \\ \nonumber 
p_\text{cold}(n,Y_p) =  p\left(n, Y_{p, \beta}, T=0\right)
+p_{\text {sym}}(n, T=0)\left[\left(1-2 Y_{p}\right)^{2}-\left(1-2 Y_{p, \beta}\right)^{2}\right] + 3K\left(Y_{p}^{4 / 3}-Y_{p, \beta}^{4 / 3}\right) n^{4 / 3} \nonumber 
\end{multline}
\end{widetext} 
where the cold symmetry energy or symmetry pressure is denoted by $E_{\rm sym}$ or $p_{\rm sym}$ and $K = (3\pi^2)^{1/3}$. The thermal terms in the smoothed approximation are provided in Eq. (\ref{Full_Energy}) and (\ref{Full_Pressure}) below.
\begin{widetext} 
\begin{align} 
    E_{th}(n,Y_p,T)=
    \frac{4\sigma f_s T^4}{n} +\biggl(\left(\frac{3T}{2}\right) ^{-1} + [a(0.5n,0.5M_{SM}^*)+a(Y_pn,m_e)Y_p] ^{-1} T^{-2}\biggr)^{-1}\,,
    \label{Full_Energy}
\end{align} 
\end{widetext} 
where $\sigma$=$\pi^2$/60 is the Stefan-Boltzmann constant, $m_e$ is the electron mass, $f_s$ is the number of ultra-relativistic species that contribute to the thermal energy (taking the value of either $1$ for $T\leq 1$ MeV or $(11/4)$ for $T > 1$ MeV provided neutrinos are neglected above $T\sim$ 10 MeV), and $a(n,M_{SM}^*)$ is a level density parameter that relies on the Dirac effective mass of the nucleon in symmetric matter, $M_{SM}^*$, and the baryon density. This parameter is defined in Eq. (\ref{a parameter}). Similarly,
\begin{multline}
    p_\text{th}(n,Y_p,T) =
    \frac{4\sigma f_s T^4}{3}\\
    \qquad + \biggr(\left({nT}\right) ^{-1}+ \biggr[\frac{\partial a(0.5n,0.5M_{SM}^*)}{\partial n}\\
    \qquad +\frac{\partial a(Y_pn,m_e)Y_p}{\partial n}\biggr] ^{-1}n^{-2}T^{-2}\biggr)^{-1} \,,
    \label{Full_Pressure}
\end{multline}

\begin{equation}
a\left(n_{q}, M^{*}_q\right) \equiv \frac{\pi^{2}}{2} \frac{\sqrt{\left(3 \pi^{2} n_{q}\right)^{2 / 3}+M^{*}_q\left(n_{q}\right)^{2}}}{\left(3 \pi^{2} n_{q}\right)^{2 / 3}}\,,
\label{a parameter}
\end{equation}

\begin{equation}
M_q^{*}\left(n_{q}\right)=\left\{\left(m c^{2}\right)^{-b}+\left[m c^{2}\left(\frac{n_{q}}{n_{0}}\right)^{-\alpha}\right]^{-b}\right\}^{-1 / b}\,,
\label{M parameter}
\end{equation}

where $n_q$ refers to the density of a single species $q$ (nucleon or lepton), $m$ is its vacuum mass and $n_0$ is a fiducial density. We chose $b=2$,$\alpha=0.87$ and $n_0=0.13$ as in~\cite{2019ApJ...875...12R} since these provide best fits to the density dependence of the common nucleonic effective mass in symmetric matter $M_{SM}^*$ for a wide variety of EOS. Eqs.(\ref{Full_Energy}), (\ref{Full_Pressure}), (\ref{a parameter}) and (\ref{M parameter}) were used for all our numerical computations involving thermal terms in the energy and pressure. 

\newpage

\bibliography{bib}
\end{document}